\documentclass[aps,prb,twocolumn,superscriptaddress,10pt]{revtex4-1}

\usepackage{amsmath,amssymb,amsfonts,upgreek}
\usepackage{color}
\usepackage{graphicx}
\usepackage{setspace}
\usepackage{bm}
\usepackage[export]{adjustbox}
\usepackage{comment}
\usepackage{hyperref}

\graphicspath{{./}}

\begin{document}

\title{Energetics of the coupled electronic-structural transition in the rare-earth nickelates}

\author{Alexander Hampel}
\email{alexander.hampel@mat.ethz.ch}
\affiliation{Materials Theory, ETH Z\"u{}rich, Wolfgang-Pauli-Strasse 27, 8093 Z\"u{}rich, Switzerland}
\author{Peitao Liu}
\email{peitao.liu@univie.ac.at}
\affiliation{Faculty of Physics, Computational Materials Physics, University of Vienna, Vienna A-1090, Austria}
\author{Cesare Franchini}
\email{cesare.franchini@univie.ac.at}
\affiliation{Faculty of Physics, Computational Materials Physics, University of Vienna, Vienna A-1090, Austria}
\affiliation{Dipartimento di Fisica e Astronomia, Universit\`{a} di Bologna, 40127 Bologna, Italy}
\author{Claude Ederer}
\email{claude.ederer@mat.ethz.ch}
\affiliation{Materials Theory, ETH Z\"u{}rich, Wolfgang-Pauli-Strasse 27, 8093 Z\"u{}rich, Switzerland}

\date{\today}

\begin{abstract}

Rare-earth nickelates exhibit a metal-insulator transition accompanied by a
structural distortion that breaks the symmetry between formerly equivalent Ni
sites. The quantitative theoretical description of this coupled
electronic-structural instability is extremely challenging. Here, we address
this issue by simultaneously taking into account both structural and electronic
degrees of freedom using a charge self-consistent combination of density
functional theory and dynamical mean-field theory, together with screened
interaction parameters obtained from the constrained random phase approximation.
Our total energy calculations show that the coupling to an electronic
instability towards a charge disproportionated insulating state is crucial to
stabilize the structural distortion, leading to a clear first order character of
the coupled transition.
The decreasing octahedral rotations across the series suppress this electronic
instability and simultaneously increase the screening of the effective Coulomb
interaction, thus weakening the correlation effects responsible for the
metal-insulator transition.
Our approach allows to obtain accurate values for the structural distortion and
thus facilitates a comprehensive understanding, both qualitatively and
quantitatively, of the complex interplay between structural properties and
electronic correlation effects across the nickelate series.

\end{abstract}

\maketitle


Complex transition metal oxides exhibit a variety of phenomena, such
as, e.g., multiferroicity~\cite{khomskii2009trend}, non-Fermi liquid
behavior~\cite{RevModPhys.73.797}, high-temperature
superconductivity~\cite{RevModPhys.66.763}, or metal-insulator
transitions~\cite{RevModPhys.70.1039}, which are not only very
intriguing, but are also of high interest for future technological
applications~\cite{Heber:2009,Takagi/Hwang:2010,Zhou_Ramanathan:2013}.
However, the quantitative predictive description of these materials
and their properties represents a major challenge for modern
computational materials science, due to the importance of electronic
correlation effects as well as due to the intimate coupling between
electronic, magnetic, and structural degrees of
freedom.~\cite{RevModPhys.70.1039,Dagotto/Tokura:2008}

An example, which has received considerable attention recently, is the
family of rare-earth nickelates, \textit{R}NiO$_3$, with $R$=La-Lu and
Y, which exhibit a rich phase diagram that is highly tunable by
strain, doping, and electromagnetic fields~\cite{Medarde:1997vt,Catalan:2008ew,catalano.review,PhysRevB.91.195138,ADMA:ADMA201003241,Middey:2016}.
All members of the nickelate series (except LaNiO$_3$) exhibit a
metal-insulator transition (MIT) as a function of temperature, which is
accompanied by a structural distortion that lowers the space group symmetry from
orthorhombic $Pbnm$, where all Ni sites are symmetry-equivalent, to monoclinic
$P2_1/n$, with two inequivalent types of Ni
sites~\cite{Alonso_et_al:1999,Alonso:1999gk,Alonso:2000fz,Alonso:2001bs}.
The structural distortion results in a three-dimensional checkerboard-like
arrangement of long bond (LB) and short bond (SB) oxygen octahedra surrounding
the two inequivalent Ni sites (see Fig.~\ref{fig:bdi-phase-csc-dmftdc}a),
and corresponds to a zone-boundary breathing mode of the
octahedral network with symmetry label
$R_1^+$~\cite{Balachandran:2013cg}.
In addition, all systems exhibit antiferromagnetic (AFM) order at low
temperatures.~\cite{Garcia-Munoz_Rodriguez-Carvajal_Lacorre:1992,Medarde:1997vt,Guo:2018}
For $R$ from Lu to Sm, the AFM transition occurs at lower temperatures
than the MIT, whereas for $R$=Nd and Pr, the magnetic transition
coincides with the MIT.  AFM order in LaNiO$_3$ was only reported
recently~\cite{Guo:2018} and is still under
discussion~\cite{Subedi:2017}. Due to challenges in synthesis,
experimental data on the bulk materials is relatively sparse,
and quantitative predictive calculations are therefore highly valuable
to gain a better understanding of the underlying mechanisms.

Different theoretical and computational approaches have highlighted
different aspects of the coupled structural-electronic transition in
the nickelates, thereby focusing either on structural or electronic
aspects~\cite{Park:2012hg,Park2014short,Park2014long,Subedi:2015en,Haule2017,Mercy2017,Varignon:2017is,hampel2017}.
Density functional theory plus Hubbard $U$ (DFT+$U$) calculations have
recently emphasized the coupling between the breathing mode and other
structural distortions such as octahedral rotations, as well as the
effect of magnetic order.~\cite{Mercy2017,hampel2017,Varignon:2017is}
However, these calculations cannot properly describe the transition from the
paramagnetic metal to the paramagnetic insulator observed in all
nickelates with $R$ cations smaller than Nd, and thus cannot correctly
capture the important electronic instability. Using DFT plus dynamical
mean field theory (DFT+DMFT)~\cite{PhysRevB.74.125120}, the MIT has
been classified as site-selective Mott transition~\cite{Park:2012hg},
where an electronic instability drives the system towards a charge-
\mbox{(or bond-)} disproportionated insulator.~\cite{Subedi:2015en}
However, the capability of DFT+DMFT to address structural properties
is currently not well established, even though promising results have
been achieved in previous
work~\cite{Park2014short,Park2014long,Haule2017}, employing either
simplified interpolation procedures between different structures,
fixing lattice parameters to experimental data, or using {\it ad hoc}
values for the interaction parameters.

Here, we combine a systematic analysis of the structural energetics,
with an accurate DFT+DMFT-based description of the electronic
structure, using screened interaction parameters obtained within the
constrained random phase approximation
(cRPA).~\cite{PhysRevB.70.195104} Our analysis thus incorporates both
structural and electronic effects, and leads to a transparent and
physically sound picture of the MIT in the nickelates, which also
allows to obtain accurate structural parameters across the whole
series. We find that the electronic instability is crucial to
stabilize the breathing mode distortion by essentially
``renormalizing'' the corresponding total energy surface, resulting in
a coupled structural-electronic first order transition. Trends across
the series are driven by the degree of octahedral
rotations,~\cite{Mercy2017} which control both the strength of the
electronic instability as well as the magnitude of the screened
interaction parameters.


\section*{Results}

\subsection*{Relaxation of $Pbnm$ structures and definition of
  correlated subspace}

All systems are fully relaxed within the high-temperature $Pbnm$ space
group using non-spinpolarized DFT calculations. We then use symmetry-based mode
decomposition~\cite{PerezMato:2010ix} to analyze the relaxed $Pbnm$
structures and quantify the amplitudes of the various distortion
modes. The mode decomposition allows for a clear conceptional
distinction between different structural degrees of freedom, which
enables us to obtain those structural degrees of freedom for which
correlation effects are not crucial from standard DFT calculations,
while the important breathing mode distortion is then obtained from
DFT+DMFT total energy calculations.
For further details on the DFT results and our distortion mode
analysis we refer to our previous work~\cite{hampel2017}.

Next, we construct a suitable low energy electronic subspace, for
which the electron-electron interaction is treated within DMFT.  Here,
we follow the ideas of
\citeauthor{Subedi:2015en}~\cite{Subedi:2015en}, and construct Wannier
functions only for a minimal set of bands with predominant Ni-$e_g$
character around the Fermi level, which in all cases (except
LaNiO$_3$) is well separated from other bands at lower and higher
energies.
The Wannier functions are then used as localized basis orbitals to
construct the effective impurity problems for our fully charge
self-consistent (CSC) DFT+DMFT calculations,~\cite{PhysRevB.76.235101}
where the LB and SB Ni sites are treated as two separate impurity
problems (even for zero $R_1^+$ amplitude) coupled through the
DFT+DMFT self-consistency loop, and the system is constrained to
remain paramagnetic.
More details on the construction of the Wannier functions and the
technical aspects of our CSC DFT+DMFT calculations can be found in the
``Methods'' section.

\subsection*{$(U,J)$ Phase diagrams}

\begin{figure}
    \includegraphics[width=0.425\textwidth]{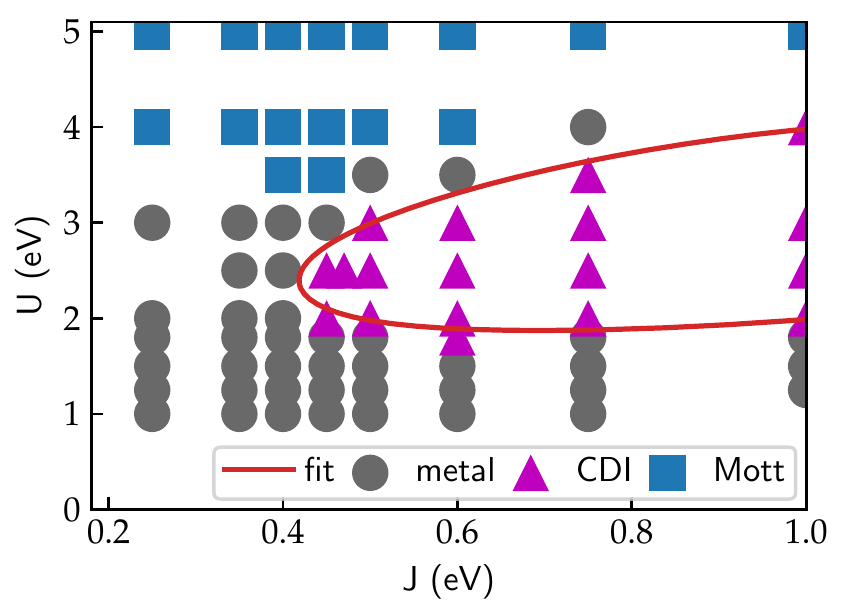}
    \caption{Phase diagram as a function of interaction parameters $U$
      and $J$ for the relaxed $Pbnm$ structure of LuNiO$_3$, i.e.,
      $R_1^+=0.0$ \r{A}. Each calculation is represented by a
      marker. Three different phases can be identified, indicated by
      different symbols: metallic (gray circles), Mott-insulator (blue
      squares), and charge-disproportionated insulator (CDI, magenta
      triangles). The boundary of the CDI phase is fitted by the red
      line.}
    \label{fig:lu_phase_diagram}
\end{figure}

\begin{figure*}
\centering
\includegraphics[width=1.0\textwidth]{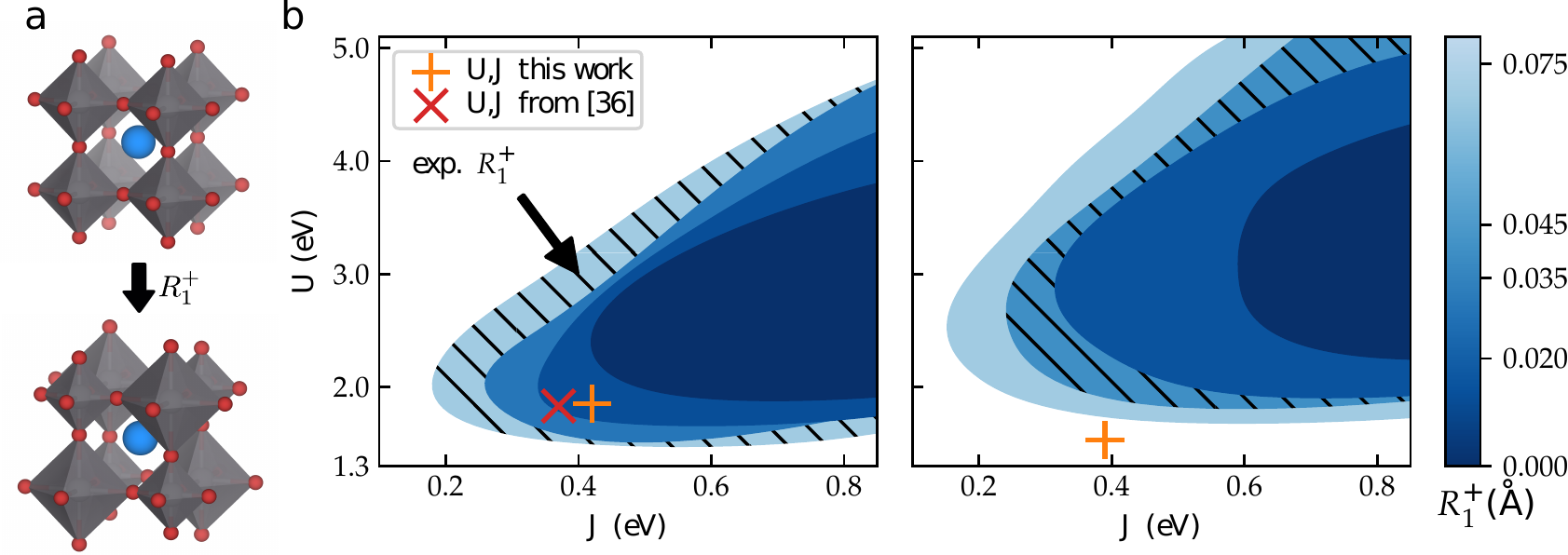}
\caption{ \textbf{a} Illustration of the $R_1^+$ breathing mode
  distortion. \textbf{b} Extension of the CDI phase within the $(U,J)$
  phase diagram for varying $R_1^+$ breathing mode amplitude for
  LuNiO$_3$ (left) and PrNiO$_3$ (right). Each $R_1^+$ amplitude is
  represented by a different brightness level, according to the color
  scale on the right, starting from $R_1^+=0.0$~\r{A} (darkest) to
  $R_1^+=0.075$~\r{A} (brightest). The levels corresponding to the
  experimental $R_1^+$ amplitudes for $R$=Lu~\cite{Alonso:2001bs} and
  $R=$Pr~\cite{Medarde2008}, respectively, are
  highlighted by diagonal stripes. The obtained cRPA values
  for $U$ and $J$ are marked by orange crosses and compared to the
  values from Ref.~\onlinecite{Seth_Georges:2017} for LuNiO$_3$ (red
  diagonal cross).  }
\label{fig:bdi-phase-csc-dmftdc}
\end{figure*}

We first establish the main overall effect of the interaction parameters $U$ and
$J$ on the electronic properties of LuNiO$_3$ within the high symmetry $Pbnm$
structure, i.e. $R_1^+=0.0$~\r{A}.
The resulting phase diagram is presented in
Fig.~\ref{fig:lu_phase_diagram}. Analogously to
Ref.~\onlinecite{Subedi:2015en}, we can identify three distinct
phases: First, a standard Mott-insulating phase for large $U$ values,
with vanishing spectral weight around the Fermi level,
$A(\omega=0)=0$, and equal occupation of all Ni sites.
Second, another insulating phase for moderate $U$ values of around
2\,eV to 3.5\,eV and relatively large $J$ ($\gtrsim 0.4$\,eV), which is
characterized by a strong difference in total occupation of the
Wannier functions centered on LB and SB Ni sites, respectively
($n_{\text{LB}} \geq 1.5$ and $n_{\text{SB}} \leq 0.5$). We denote this phase as
charge disproportionated insulating (CDI) phase~\cite{Mazin:2007}.
Third, a metallic phase for small $U$ values in between the two insulating
regions, with equal occupation on all Ni sites, $n_{SB} \approx n_{LB} \approx
1.0$, and non-vanishing spectral weight at the Fermi level, $A(\omega=0)>0$.

The CDI phase has been identified as the insulating low-temperature phase of
nickelates in Ref.~\onlinecite{Subedi:2015en}, where it has also been shown that
the strong charge disproportionation is linked to the MIT (in
Ref.~\onlinecite{Subedi:2015en} this phase has been termed ``bond
disproportionated insulating'').
We note that the Wannier basis within our low energy subspace, while
being centered on the Ni sites with strong $e_g$ character, also exhibits strong
tails on the O ligands, and thus the corresponding charge is distributed over
the central Ni atom and the surrounding O atoms.
The strong charge disproportionation found within our chosen basis set
is thus fully consistent with the observation that the integrated
charge around the two different Ni atoms differs only
marginally~\cite{Park:2012hg}.
Alternatively, within a negative charge transfer picture, the MIT can also be
described, using a more atomic-like basis, as $(d^8\underline{L})_{i} \
(d^8\underline{L})_{j} \rightarrow (d^8\underline{L}^2)_{\text{SB}} \
(d^8)_\text{LB}$, where $\underline{L}$ denotes a ligand hole ({\it c.f.}
Refs.~\onlinecite{Park:2012hg,Johnston_Sawatzky:2014,Varignon:2017is,mandal_franchini:2017}).

One should also note that the CDI phase appears even though all Ni
sites are structurally equivalent ($R_1^+=0$ in
Fig.~\ref{fig:lu_phase_diagram}), which indicates an electronic
instability towards spontaneous charge disproportionation. This has
already been found in Ref.~\onlinecite{Subedi:2015en}, and indicates
that a purely lattice-based description is incomplete.
Moreover, within our CSC DFT+DMFT calculations, the CDI phase appears at
significantly lower $J$ and a more confined $U$ range compared to the non-CSC
calculations of Ref.~\onlinecite{Subedi:2015en}. A similar reduction of $J$
values necessary to stabilize the CDI phase has also been achieved in the
non-CSC DFT+DMFT calculations of Ref.~\onlinecite{Seth_Georges:2017}, through
the introduction of an (effective) inter-site Hartree interaction.This suggests
that the latter can indeed mimic the main effect of a CSC calculation, where the
charge density, and thus the local occupations, are updated and the Hartree
energy is recalculated in each CSC step.

Next, we investigate how the electronic instability corresponding to the CDI
phase couples to the structural $R_1^+$ breathing mode distortion. For this, we
vary only the $R_1^+$ amplitude, while keeping all other structural parameters
fixed to the fully relaxed (within nonmagnetic DFT) $Pbnm$ structures, and
calculate $(U,J)$ phase diagrams for different values of the $R_1^+$ amplitude.
We do this for both LuNiO$_3$ and PrNiO$_3$, i.e., for the two compounds with
the smallest and largest rare earth cations within the series that exhibit the
MIT.
The $(U,J)$ range of the CDI phase for a given $R_1^+$ amplitude is
then extracted by interpolating the convex hull of the phase boundary
(similar to the red line in Fig.~\ref{fig:lu_phase_diagram}). The
results are summarized in Fig.~\ref{fig:bdi-phase-csc-dmftdc}b.

In both cases, $R$=Lu and $R$=Pr, the $R_1^+$ amplitude couples
strongly to the CDI state, and increases the corresponding area within
the $(U,J)$ phase diagam.  In particular, the minimal $J$ required to
stabilize the CDI phase is significantly lowered. Furthermore, also
for $R$=Pr, there is a spontaneous instability towards the formation
of a CDI state, but the corresponding $(U,J)$ range is noticeably
smaller than for $R$=Lu.
In addition, the minimal $U$ required to stabilize the CDI phase for a given
$R_1^+$ amplitude is slighty higher for $R$=Pr than for $R$=Lu. We note that,
since the $R$ ions do not contribute noticeably to any electronic states close
to the Fermi level, the differences between the two materials are mainly due to
the different underlying $Pbnm$ structures, specifically the weaker octahedral
tilts in PrNiO$_3$ compared to LuNiO$_3$. This increases the electronic
bandwidth, which opposes the tendency towards charge disproportionation.

\subsection*{Calculation of interaction parameters}

\begin{figure}[t]
    \centering
    \includegraphics[width=0.45\textwidth]{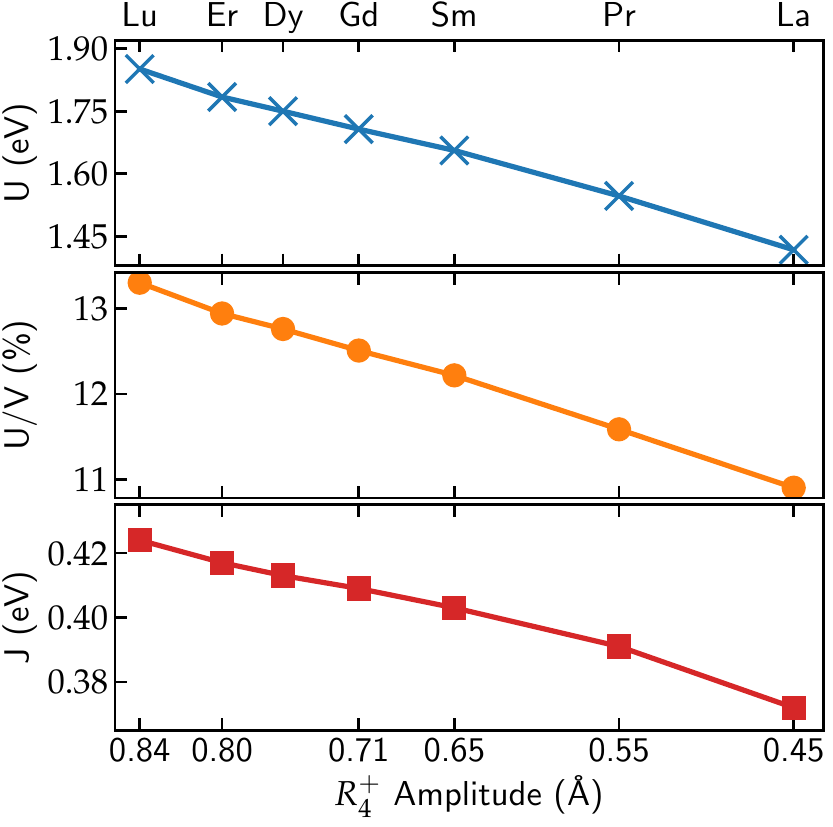}
    \caption{Screened onsite Hubbard-Kanamori interaction parameters
      $U$ (top) and $J$ (bottom) for the $e_g$ orbitals within our
      low-energy subspace across the nickelate series as a function of
      the octahedral tilt amplitude $R_4^+$. Additionally,
      the ratio between $U$ and the corresponding bare (unscreened)
      interaction parameter $V$ is shown (middle).}
    \label{fig:crpa_r1}
\end{figure}

So far we have varied $U$ and $J$ in order to obtain the general
structure of the phase diagram. Next, we calculate $U$ and $J$
corresponding to our correlated subspace for all systems across the
series to see where in these phase diagrams the real materials are
located.
We use cRPA~\cite{PhysRevB.70.195104} to extract the partially screened
interaction parameters $(U,J)$ within the Hubbard-Kanamori
parameterization, by separating off the screening channels related to
electronic transitions within the correlated $e_g$ subspace from all
other transitions (see also Methods section).

The results of these cRPA calculations are shown in
Fig.~\ref{fig:crpa_r1} as a function of the $R$ cation and the
corresponding $R_4^+$ amplitude, i.e., the main octahedral tilt mode
in the $Pbnm$ structure. The effective interaction parameters $U$
corresponding to our $e_g$ correlated subspace are strongly screened
compared to the bare interaction parameters $V$. For LuNiO$_3$, we
obtain $V=13.91$~eV and $U=1.85$~eV, while $J=0.42$~eV with a
corresponding bare value of 0.65~eV. This is in good agreement with
Ref.~\onlinecite{Seth_Georges:2017}, which obtained $U=1.83$~eV and
$J=0.37$~eV using the experimental $P2_1/n$ structure. Furthermore,
both $U$ and $J$ decrease monotonically across the series (for
decreasing $R_4^+$ amplitude), leading to an additional reduction of
$U$ by 25\% in LaNiO$_3$ compared to LuNiO$_3$. This decrease is also
observed in the ratio $U/V$, indicating that it is due to an even
stronger screening for $R$=La compared to $R$=Lu.

Our calculated $(U,J)$ parameters for $R$=Lu and $R$=Pr are also
marked in the corresponding phase diagrams in
Fig.~\ref{fig:bdi-phase-csc-dmftdc}. It is apparent,
that for $R$=Lu the calculated cRPA values are well within the
stability region of the CDI phase, even for a relatively small $R_1^+$
amplitude of $0.02$\,\r{A}. In contrast, for $R$=Pr, the values are
outside the CDI phase even for $R_1^+$ amplitudes larger than the one
experimentally observed. Thus, at their respective experimental
breathing mode amplitudes, our calculations predict a paramagnetic CDI
state for LuNiO$_3$ but not for PrNiO$_3$.

\subsection*{Lattice energetics}

\begin{figure}[t]
    \centering
    \includegraphics[width=0.45\textwidth]{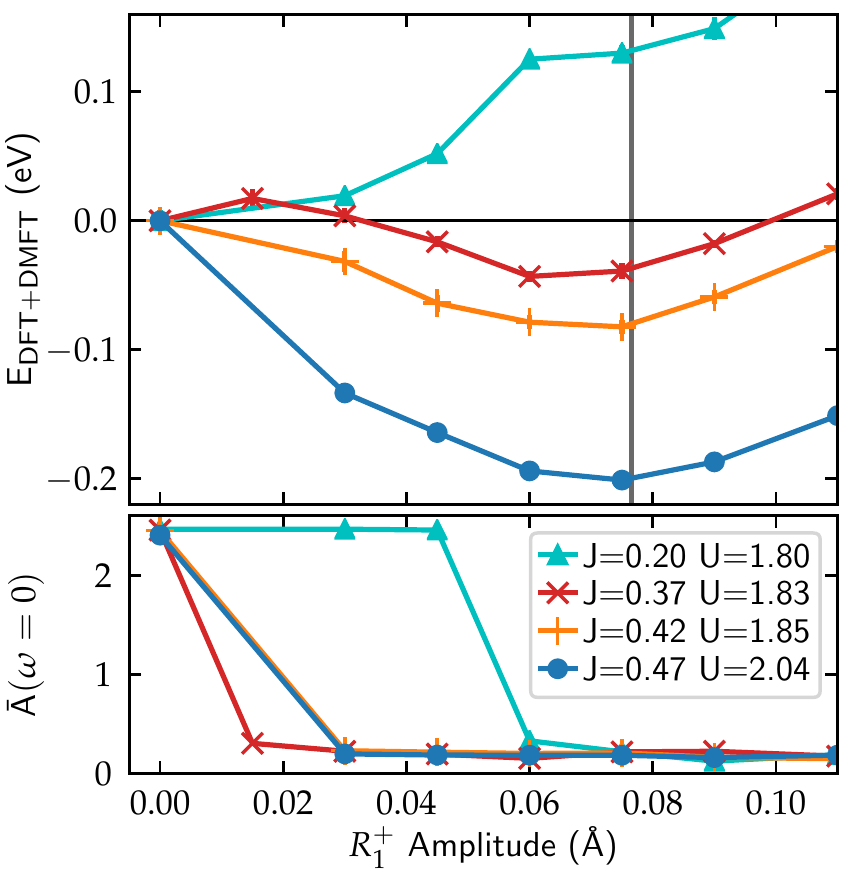}
    \caption{Top: Total energy, $E_{\text{DFT+DMFT}}$, as a function of
      the $R_1^+$ breathing mode amplitude for LuNiO$_3$ using
      different values for the interaction parameters $U$ and $J$. The
      experimental amplitude  ($R_1^+=0.075$~\r{A}~\cite{Alonso:2001bs}) is marked by
      the gray vertical line. Bottom: Corresponding spectral weight at
      the Fermi level, indicating the MIT as a function of $R_1^+$
      amplitude.}
    \label{fig:e_tot_lunio3}
\end{figure}

\begin{figure}[t]
    \centering
    \includegraphics[width=0.45\textwidth]{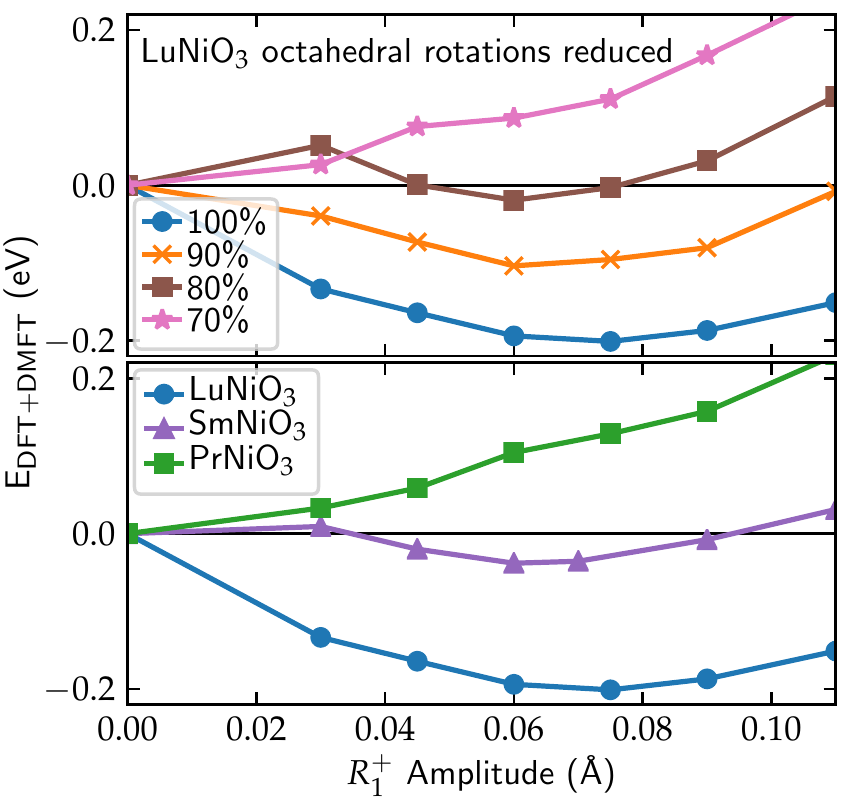}
    \caption{Top: Total energy as a function of the $R_1^+$ breathing
      mode amplitude for LuNiO$_3$ with octahedral rotation amplitudes
      reduced to 90\,\%, 80\,\%, and
      70\,\% (for $U=2.04$\,eV and $J=0.47$\,eV). Bottom:
      Corresponding data for various materials across the nickelate
      series. Here, $(U,J)$ values are increased by 10\% compared to
      the results of the cRPA calculations ($U=2.04$\,eV/$J=0.47$\,eV
      for LuNiO$_3$, $U=1.82$\,eV/$J=0.44$\,eV for SmNiO$_3$, and
      $U=1.70$\,eV/$J=0.43$\,eV for PrNiO$_3$).
        }
    \label{fig:e_tot_csc_crpa}
\end{figure}

Up to now, we have been addressing the stability of the CDI phase for a given
(fixed) $R_1^+$ amplitude. Now, we will address the stability of the $R_1^+$
mode itself and calculate its amplitude across the series using total energy
calculations within CSC DFT+DMFT. The symmetry-based mode decomposition
allows us to systematically vary only the $R_1^+$ mode, while keeping all other
structural parameters fixed to the values obtained from the nonmagnetic DFT
calculations.
Thus, in contrast to interpolation procedures as in
Refs.~\onlinecite{Park2014long} or \onlinecite{Haule2017}, our
approach excludes any additional energy contributions related to
simultaneous changes in other structural distortions, in particular
the octahedral tilt modes.

Fig.~\ref{fig:e_tot_lunio3} shows the total energy and the spectral
weight around the Fermi level, $\bar{A}(\omega=0)$, as a function of the
$R_1^+$ amplitude for LuNiO$_3$, calculated using different values for
$(U,J)$. First, we focus on the results obtained using our cRPA
calculated values ($J=0.42$\,eV, $U=1.85$\,eV, orange crosses). It can
be seen, that the energy indeed exhibits a minimum for an $R_1^+$
amplitude very close to the experimental value. Furthermore, as seen
from $\bar{A}(\omega=0)$, the system undergoes a MIT for
increasing $R_1^+$ amplitude and is clearly insulating in the region
around the energy minimum. Thus, our CSC DFT+DMFT calculations
together with the calculated cRPA interaction parameters correctly
predict the CDI ground state for LuNiO$_3$, and furthermore result in
a breathing mode amplitude that is in excellent agreement with
experimental data.

To see how subtle changes in $(U,J)$ influence the energetics of the
system, we also perform calculations using the cRPA values obtained in
Ref.~\onlinecite{Seth_Georges:2017} ($J=0.37$\,eV, $U=1.83$\,eV, red
diagonal crosses). In this case, we obtain a more shallow energy
minimum at a slightly reduced amplitude of $R_1^+=0.06$~\r{A}. This
reduction is mainly caused by the slightly smaller $J$.
Moving the values of $(U,J)$ even closer to the boundary of the stability region
of the CDI phase for the experimental $R_1^+$ amplitude, {\it cf}.
Fig.~\ref{fig:bdi-phase-csc-dmftdc} (e.g., $J=0.2$\,eV, $U=1.8$\,eV, cyan
triangles), results in a loss of the energy minimum for finite $R_1^+$
amplitude. Nevertheless, a kink in the total energy is clearly visible at the
$R_1^+$ amplitude for which the system becomes insulating, indicating the strong
coupling between the structural distortion and the MIT. A similar kink can also
be recognized (for rather small $R_1^+$ amplitude) in the total energy obtained
for $J=0.37$\,eV and $U=$1.83\,eV, resulting in an additional local energy
minimum at $R_1^+=0$, a typical hallmark of a first order structural transition.
In addition, we also perform calculations where $(U,J)$ are increased by 10\,\%
compared to our cRPA values ($J=0.47$\,eV, $U=2.04$\,eV, red circles), which
leads to a deeper energy minimum and an $R_1^+$ amplitude in near perfect
agreement with experiment.

Next, we investigate the influence of the octahedral rotations on the
energetics of the $R_1^+$ mode, where we perform a series of
calculations for LuNiO$_3$ with artificially decreased octahedral
rotations (see methods section), fixed $(U, J)$, and fixed volume. As
can be seen from the data shown in the top panel of
Fig.~\ref{fig:e_tot_csc_crpa}, decreasing the amplitude of the
octahedral rotations to 70\,\%, which corresponds roughly to the
amplitudes found for PrNiO$_3$, leads to a vanishing of the minimum at
non-zero $R_1^+$ amplitude. This confirms that the reduction of the
octahedral rotation amplitudes plays a crucial role in the energetics
of the breathing mode distortion and in determining the trend across
the nickelate series.

Finally, we examine how the energetics of the $R_1^+$ mode varies
across the series, by comparing the two end members LuNiO$_3$ and
PrNiO$_3$, as well as SmNiO$_3$, which is the compound with the
largest $R$ cation in the series that still exhibits a paramagnetic
CDI state. In each case we use $(U,J)$ values that are increased by
10\,\% relative to the corresponding cRPA values.  The use of such
slightly increased interaction parameters is motivated by the
observation that the $U$ values obtained from the static limit of the
(frequency-dependent) screened cRPA interaction are often too small to
reproduce experimental data for various
materials~\cite{Casula:2012,PhysRevB.74.125120,anisimov2009,roekeghem_biermann:2014}.
The results are depicted in Fig.~\ref{fig:e_tot_csc_crpa}.

As discussed above, for LuNiO$_3$ (blue circles), we obtain an
energy minimum exactly at the experimentally observed amplitude.
For SmNiO$_3$ (purple triangles), we obtain a much more shallow minimum at
$R_1^+=0.06$~\r{A}, which corresponds to a reduction by $\approx 20$\,\%
compared to LuNiO$_3$. Unfortunately, structural refinements for SmNiO$_3$ are
only available within the $Pbnm$ space group, and thus no information on the
$R_1^+$ amplitude exists~\cite{Rodrigues_Medarde:1998}. However, the reduction
of the $R_1^+$ amplitude from $R$=Lu to $R$=Sm is much more pronounced compared
to previous DFT+$U$ calculations with AFM order~\cite{hampel2017}, where the
reduction is only about $8$\,\%.

For PrNiO$_3$ (green squares), no stable $R_1^+$ amplitude is obtained
within our paramagnetic DFT+DMFT calculations, but a kink marking the
MIT is still visible at $R_1^+=0.06$~\r{A}. This is also in agreement
with the experimental observation that no paramagnetic CDI phase
occurs in PrNiO$_3$~\cite{Medarde:1997vt}.
Furthermore, it was recently demonstrated using DFT+DMFT calculations that for
NdNiO$_3$ the CDI state becomes only favorable in the antiferromagnetically
ordered state~\cite{Haule2017}. Our results indicate that this also holds for
PrNiO$_3$, while in SmNiO$_3$ a stable $R_1^+$ amplitude can be found even in the
paramagnetic case. Thus, the phase boundaries across the series are correctly
described within the DFT+DMFT approach.
We further note that, considering the $(U,J)$ phase diagrams for PrNiO$_3$ in
Fig.~\ref{fig:bdi-phase-csc-dmftdc}, a $U$ of up to 2.5 or even 3\,eV would be
required to put PrNiO$_3$ well within the CDI phase region at its experimental
$R_1^+$ amplitude, which appears necessary to obtain a stable $R_1^+$ amplitude.
However, such a large $U$ seems highly unrealistic considering the calculated
cRPA values.

\section*{Discussion}

In summary, the successful application of CSC DFT+DMFT and
symmetry-based mode analysis, without {\it ad hoc} assumptions
regarding the strength of the Hubbard interaction or fixing structural
parameters to experimental data, allows to elucidate the nature of the
coupled electronic-structural transition across the nickelate series.
Our analysis reveals that the MIT, which is related to an electronic
instability towards spontaneous charge disproportionation, leads
to a significant restructuring of the energy landscape, indicated by a
kink in the calculated total energy. This creates a minimum at a
finite $R_1^+$ amplitude (for appropriate $U$ and $J$), and suggests a
first order character of the coupled structural and electronic
transition in the PM case, in agreement with experimental
observations~\cite{catalano.review} for both
SmNiO$_3$~\cite{Cacho:1999} and YNiO$_3$.~\cite{Alonso_et_al:1999} We
note that, since a certain critical value of $R_1^+$ is necessary to
induce the MIT (see, e.g., Fig.~\ref{fig:e_tot_lunio3}), a second order
structural transition would imply the existence of an intermediate
structurally distorted metallic phase, inconsistent with experimental
observations.

The strength of the electronic instability towards spontaneous charge
disproportionation and thus the stability range of the CDI phase, is strongly
affected by the amplitude of the octahedral rotations, varying across the
series.
This is in agreement with Ref.~\onlinecite{Mercy2017}, but in addition we
show that to arrive at a fully coherent picture, with correct phase
boundaries, it is crucial to treat both electronic and structural
degrees of freedom on equal footing.
For example, even though a CDI state can be obtained for
PrNiO$_3$ for fixed $R_1^+$ amplitude $>0.06$\,\r{A}, our calculations
show that this is indeed energetically unstable.
In addition, the octahedral rotations also influence the screening of
the effective interaction parameters, disfavoring the CDI state for
larger $R$ cations.
As a result, magnetic order appears to be crucial to stabilize the
breathing mode distortion for both $R$=Nd and Pr.

Moreover, our calculations not only lead to a coherent picture of the
MIT, but also allow to obtain accurate structural parameters across
the nickelate series. Furthermore, this is achieved using only a
minimal correlated subspace.
We note that the use of such a reduced correlated subspace can be
advantageous, since it not only allows to reduce the computational
effort (due to less degrees of freedom), but also because the double
counting problem is typically less severe if the O-$p$ dominated bands
are not included in the energy window of the correlated
subspace.~\cite{Janson_Held:2018,Karolak_Lichtenstein:2010} In the
present case, the resulting more extended Wannier functions, which
also incorporate the hybridization with the surrounding ligands, also
provide a rather intuitive picture of the underlying charge
disproportionation.

Finally, our study represents the successful application of a
combination of several state-of-the-art methods that allows to tackle
other open issues related to the entanglement of structural and
electronic properties in correlated materials, such as Jahn-Teller and
Peierls instabilities, charge density wave, or polarons.

\section*{Methods}

\paragraph*{DFT calculations}

All DFT calculations are performed using the projector augmented wave (PAW)
method~\cite{Blochl:1994dx} implemented in the ``Vienna Ab initio Simulation
Package''(VASP)~\cite{Kresse:1993bz,Kresse:1996kl,Kresse:1999dk} and the
exchange correlation functional according to Perdew, Burke, and
Ernzerhof~\cite{Perdew:1996iq}. For Ni, the 3$p$ semi-core states are included
as valence electrons in the PAW potential. For the rare-earth atoms, we use PAW
potentials corresponding to a $3+$ valence state with $f$-electrons frozen into
the core and, depending on the rare-earth cation, the corresponding $5p$ and
$5s$ states are also included as valence electrons.
A $k$-point mesh with $10 \times 10 \times 8$ grid points along the three
reciprocal lattice directions is used and a plane wave energy cut-off of 550~eV
is chosen for the 20 atom $Pbnm$ unit cell.
All structures are fully relaxed, both internal parameters and lattice
parameters, until the forces acting on all atoms are smaller than $10^{-4}$
eV/\r{A}.
As in Ref.~\onlinecite{hampel2017}, we perform calculations for LaNiO$_3$ within
the $Pbnm$ and $P2_1/n$ space groups, to allow for a more consistent comparison
with the rest of the series, even though LaNiO$_3$ is experimentally found in a
different space group ($R\bar{3}c$). See also the discussion in
Ref.~\onlinecite{Subedi:2017}.

\paragraph*{Distortion mode analysis}

For the symmetry-based mode decomposition~\cite{PerezMato:2010ix} we
use the software ISODISTORT \cite{Campbell:2006}. Thereby, the atomic
positions within a distorted low-symmetry crystal structure,
$\vec{r}_i^{\ \text{dist}}$, are written in terms of the positions in
a corresponding non-distorted high-symmetry reference structure,
$\vec{r}_i^{\ 0}$, plus a certain number of independent distortion
modes, described by orthonormal displacement vectors, $\vec{d}_{im}$,
and corresponding amplitudes, $A_{m}$:
\begin{align}
\vec{r}_i^{\ \text{dist}} = \vec{r}_i^{\ 0} + \sum\limits_m A_m \
\vec{d}_{im} \qquad .
\end{align}

The distortion modes of main interest here are the out-of-phase and
in-phase tilts of the oxygen octahedra, $R_4^+$ and $M_3^+$, for
characterization of the high-temperature $Pbnm$ structure, and the
$R_1^+$ breathing mode distortion within the low-temperature $P2_1/n$
structure.
A more detailed description for nickelates can be found,
e.g., in Refs.~\onlinecite{Balachandran:2013cg,hampel2017}.
For the calculations with reduced octahedral rotation amplitudes shown
in Fig.~\ref{fig:e_tot_csc_crpa}, both $R_4^+$ and $M_3^+$ modes, as
well as the $X_5^+$ mode intimately coupled to these two modes, have
been reduced by a common factor.

\paragraph*{DMFT calculations}

The Wannier functions for our CSC DFT+DMFT calculations are
constructed via projections on local Ni $e_g$ orbitals as described in
Ref.~\onlinecite{Schuler_Aichhorn:2018,PhysRevB.77.205112}, using the
\textsc{TRIQS/DFTTools} software
package.~\cite{aichhorn_dfttools_2016,parcollet_triqs_2015}
The effective impurity problems within the DMFT loop are solved with
the \textsc{TRIQS/cthyb} continuous-time hybridization-expansion
solver~\cite{Seth2016274}, including all off-diagonal spin-flip and
pair-hopping terms of the interacting Hubbard-Kanamori
Hamiltonian.~\cite{vaugier2012} The LB and SB Ni sites are treated as two
separate impurity problems (even for zero $R_1^+$ amplitude), where
the number of electrons per two Ni sites is fixed to 2, but the
occupation of each individual Ni site can vary during the calculation
(while the solution is constrained to remain paramagnetic).

The fully-localized limit~\cite{anisimov1997} is used to correct for
the double-counting (DC) in the parametrization given in
Ref.~\onlinecite{held2007}:
\begin{align}
  \label{eq:dcimp}
  \Sigma_{dc,\alpha}^{imp} = \bar{U}
  (n_{\alpha}-\frac{1}{2}) \quad ,
\end{align}
where $n_{\alpha}$ is the occupation of Ni site $\alpha$, obtained in
the DMFT loop, and the averaged Coulomb interaction is defined as
$\bar{U}=(3U-5J)/3$. Note, that in our Wannier basis the occupations
change quite drastically from the original DFT occupations and the
choice of the DC flavor can therefore influence the outcome. However,
with respect to the lattice energetics we found no difference in the
physics of the system when changing the DC scheme or using fixed DFT
occupation numbers for the calculation of the DC correction. If the
DFT occupations are used instead of the DMFT occupations, larger
interaction parameters are required to obtain the same predicted
$R_1^+$ amplitude. However, we note that the DFT occupations have no
clear physical meaning within CSC DFT+DMFT.

The spectral weight around the Fermi level, $\bar{A}(\omega=0)$, is
obtained from the imaginary time Green's
function:~\cite{PhysRevB.83.235113}
\begin{align}
    \bar{A}(\omega=0) = - \frac{\beta}{\pi} G_\text{imp} \left( \frac{\beta}{2} \right) \quad .
\end{align}
For $T=0$ ($\beta \rightarrow \infty$), $\bar{A}$ is identical to the spectral
function at $\omega=0$. For finite temperatures, it represents a weighted
average around $\omega=0$ with a width of $\sim
k_\text{B}T$~\cite{PhysRevB.83.235113}.

The total energy is calculated as described in Ref.~\onlinecite{PhysRevB.74.125120}:
\begin{align}
\begin{split}
    E_{\text{DFT+DMFT}} &= E_{\text{DFT}}[\rho] \\
    &- \frac{1}{N_k} \sum_{\lambda,\vec{k}} \epsilon_{\lambda,\vec{k}}^{\text{KS}} \ f_{\lambda \vec{k}} + \langle H_{\text{KS}} \rangle_{\text{DMFT}} \\
    & + \langle H_{\text{int}} \rangle_{\text{DMFT}} - E_{\text{DC}}^{imp} \quad .
\end{split}
\label{eq:dmft-dft-tot-en-2}
\end{align}

The first term is the DFT total energy, the second term subtracts the
band energy of the Ni $e_g$ dominated bands (index $\lambda$), the
third term evaluates the kinetic energy within the correlated subspace
via the lattice Green's function, the fourth term adds the interaction
energy, where we use the Galitskii-Migdal
formula~\cite{abrikosov2012methods,galitskii1958}, and the last term
subtracts the DC energy. To ensure good accuracy of the total energy,
we represent both $G_\text{imp}$ and $\Sigma_\text{imp}$ in the
Legendre basis~\cite{PhysRevB.84.075145} and obtain thus smooth
high-frequency tails and consistent Hartree shifts. Moreover, we
sample the total energy over a minimum of additional 60 converged DMFT
iterations after the CSC DFT+DMFT loop is converged. Convergence is
reached when the standard error of the Ni site occupation of the last
10 DFT+DMFT loops is smaller than $1.5 \times 10^{-3}$. That way we
achieve an accuracy in the total energy of $<5$\,meV. All DMFT
calculation are performed for $\beta=40$ eV$^{-1}$, which corresponds
to a temperature of 290~K.

\paragraph*{cRPA calculations}

We use the cRPA method as implemented in the VASP code~\cite{Merzuk2015}
to extract interaction parameters for our correlated subspace. These
calculations are done for the relaxed $Pbnm$
structures~\cite{hampel2017}. We follow the ideas given in the paper
of Ref.~\onlinecite{Subedi:2015en} and construct
maximally localized Wannier functions (MLWFs) for the Ni-$e_g$
dominated bands around the Fermi level using the \textsc{wannier90}
package~\cite{wannier90}. Since the corresponding bands are isolated
from other bands at higher and lower energies, no disentanglement
procedure is needed, except for LaNiO$_3$, for which we ensured that
the resulting Wannier functions are well converged and have a very
similar spread as for all other compounds of the series.

We divide the total polarization, $P$, into a contribution involving only
transitions within the effective ``$e_g$'' correlated subspace and the rest,
$P=P_{e_g} + P_r$. The constrained polarization, $P_r$, and the static limit of
the screened interaction matrix, $W_r(\omega=0)=V[1-V P_r(\omega=0)]^{-1}$, where $V$
is the bare interaction, are then calculated using a $5 \times 5 \times 3$
$k$-point mesh, a plane wave energy cut-off of $E_\text{cut}=600$\,eV, and 576
bands. Effective values for the Hubbard-Kanamori interaction parameters $(U,J)$
are extracted from $W_r(\omega=0)$ as described in
Ref.~\onlinecite{vaugier2012}. Our procedure is analogous to the calculation of
effective interaction parameters for LuNiO$_3$ in
Ref.~\onlinecite{Seth_Georges:2017}.

It should be noted that the MLWFs used for the cRPA
calculations are not completely identical to the projected Wannier
functions used as basis for the correlated subspace within our DMFT
calculations. However, test calculations for the case of LuNiO$_3$
showed only minor differences between the hopping parameters
corresponding to the MLWFs and the ones corresponding to the Wannier
functions generated by the projection scheme implemented in
VASP. Furthermore, we did not find a noticeable difference between the
screened $(U,J)$ values calculated for the MLWFs and the ones
calculated for the initial guesses for these Wannier functions, i.e.,
before the spread minimization, which are also defined from
orthogonalized projections on atomic-like orbitals. We thus conclude
that the two sets of Wannier functions are indeed very similar, and
that the cRPA values of $(U,J)$ obtained for the MLWFs are also
representative for the Wannier basis used in our DMFT calculations.

Additionally, we point out that, in contrast to what was found in
Ref.~\onlinecite{Seth_Georges:2017}, we observe only negligible
differences in the interaction parameters obtained for the relaxed
$Pbnm$ structure and the ones obtained for the experimental
low-temperature $P2_1/n$ structure for LuNiO$_3$ (1.827\,eV and
1.876\,eV compared to 1.849\,eV within $Pbnm$).
In particular, the difference of the interaction parameters on the two
inequivalent Ni sites in the $P2_1/n$ structure ($\pm$0.03\,eV) are
very small compared to the changes stemming from different degrees of
octahedral rotations (i.e., different $R$ cations), justifying the use
of constant interaction parameters for different $R_1^+$ amplitudes.
Furthermore, the differences in the intra-orbital $U$ matrix elements
between the $d_{z^2}$ and the $d_{x^2-y^2}$ orbitals are negligible
small, $\sim 0.01$\,eV, in our calculations. Therefore, all the values
of the interaction parameters are averaged over both $e_g$ orbitals.
\section*{Data Availability Statement}
The data that support the findings of this study are available from the corresponding
author upon reasonable request.
\begin{acknowledgments}

We are indebted to Oleg Peil and Antoine Georges for helpful
discussions. This work was supported by ETH Zurich and the Swiss
National Science Foundation through grant No. 200021-143265 and
through NCCR-MARVEL. Calculations have been performed on the clusters
``M\"o{}nch'' and ``Piz Daint'', both hosted by the Swiss National
Supercomputing Centre, and the ``Euler'' cluster of ETH
Zurich.

\end{acknowledgments}
\section*{Competing interests}

The Authors declare no Competing Financial or Non-Financial Interests.
\section*{Author contributions}

A.H. performed and analyzed all DFT and DMFT calculations. The cRPA
calculations were done by A.H. with the help of P.L. and supervised by
C.F. The whole project was initiated by C.E. The initial manuscript
was written by A.H. and C.E. All authors discussed the results at
different stages of the work and contributed to the final manuscript.

\bibliography{bibfile}

\end{document}